# Beyond Completion Time: A Multimodal Approach to Characterizing Eye-Hand Coordination During the Nine-Hole Peg Test in Multiple Sclerosis

Mahya Beheshti[1,2*], Rajvardhan Gadde[1,2*], Diana Maloku[4], Danae S. Manta[4], Bethany Lum[1], Silvana L. Costa[4], Todd E. Hudson[1,3], John-Ross Rizzo[1,2,3]

## Abstract

Multiple sclerosis (MS) is a chronic neurodegenerative disease associated with motor, sensory, cognitive, and visual deficits that can impair upper extremity function. Although the Nine-Hole Peg Test (9-HPT) is widely used to assess fine motor dexterity in MS, conventional scoring captures only overall completion time, providing limited insight into the eye–hand coordination processes underlying performance. This study characterized eye–hand coordination during peg placement using simultaneous eye tracking, markerless hand tracking, and object detection. Nine people with MS (PwMS) and nine healthy controls completed the 9-HPT with both hands. Gaze, hand, and target-hole positions were mapped frame by frame to pegboard coordinates. Each peg cycle was divided into object transport, during which the peg was moved from the collection bin to the target grid, object manipulation, during which the peg was aligned and inserted into the target hole, and the return, from the seated peg to the grasp of the next peg in the bin. Manipulation duration was measured from hand arrival at the target hole to completion of insertion. PwMS had longer overall completion times than controls (MS vs. controls: dominant hand, 25.9 vs. 20.7 s, $p = 0.025$; non-dominant hand, 29.5 vs. 21.5 s, $p = 0.034$). Transport duration was longer in PwMS with the non-dominant hand (563 vs. 430 ms, $p = 0.027$) but not with the dominant hand (429 vs. 402 ms, $p = 0.52$). Manipulation duration was longer in PwMS with the dominant hand (544 vs. 439 ms, $p = 0.036$) and numerically longer with the non-dominant hand (765 vs. 477 ms, $p = 0.105$). The return from the seated peg to the grasp of the next peg was longer in PwMS with both hands (dominant hand, 1028 vs. 676 ms, $p = 0.003$; non-dominant hand, 1068 vs. 700 ms, $p = 0.021$), the largest of the three phase differences. PwMS made approximately one additional fixation per peg and exhibited greater peg-to-peg variability in gaze landing position during dominant-hand performance (21.8 vs. 16.1 mm, $p = 0.027$). Saccadic amplitude and mean placement error did not differ significantly between groups, and eye movement generally preceded hand movement in both groups. These exploratory findings suggest that slowing in MS was concentrated in the phases requiring precise manipulation, the insertion of the peg and the return to the bin to pick up the next peg. Multimodal assessment of the standard 9-HPT can distinguish transport from manipulation and reveal coordination features that overall completion time alone cannot capture.

[1] Department of Rehabilitation Medicine, NYU Langone Health, 244 E38th St., New York, NY, 10016, USA.
[2] Department of Mechanical & Aerospace Engineering, NYU Tandon School of Engineering, New York, NY, 10010, USA.
[3] Department of Neurology, NYU Langone Health, 222 E40th St., New York, NY, 10016, USA.
[4] Neuropsychology & Neuroscience Laboratory, Kessler Foundation, West Orange, NJ, USA
*Authors contributed equally

## 1 Introduction

Multiple sclerosis (MS) is a chronic neurodegenerative disease of the central nervous system characterized by demyelination and axonal degeneration, with associated motor, sensory, cognitive, and visual deficits[1]. It affects approximately 2.8 million people worldwide and is a major cause of non-traumatic neurological disability in young adults [2]. Upper extremity dysfunction affects up to 76% of people with MS (PwMS), compromising grasping, fine finger control, and object manipulation, with substantial consequences for daily activities, independence, and quality of life[3]. However, the sensorimotor processes contributing to these functional difficulties remain incompletely understood.

Successful object manipulation requires coordinated eye and hand movements. Gaze typically shifts toward a task-relevant target before the hand arrives, providing spatial information to support movement planning and subsequent adjustments[4, 5]. These demands change as action progresses: transporting an object toward a target requires relatively rapid movement, whereas aligning and inserting it into a small opening requires precise fingertip control under visual and tactile guidance. In MS, disruption of the pathways supporting these processes may contribute to delayed movement initiation, corrective adjustments, and inaccuracies near the end of a movement[6, 7]. Examining eye and hand behavior across these distinct phases may therefore help identify where performance becomes impaired.

The Nine-Hole Peg Test (9-HPT) is a widely used measure of fine motor dexterity in MS and a core component of the Multiple Sclerosis Functional Composite[8]. Conventional scoring records the total time required to place nine pegs into a board and then remove them. Although clinically useful, this aggregate measure does not distinguish the movement components contributing to slower performance. During peg placement, each action can be divided into object transport, in which the peg is moved from the collection bin to the target grid, and object manipulation, in which the peg is aligned and inserted into the target hole. Separating these phases could distinguish slowing during transport from difficulty with the precise manipulation required for insertion.

Recent approaches to instrumenting the 9-HPT have incorporated hand kinematics to quantify movement velocity, smoothness, and peg-placement timing[9]. These measures provide greater detail than completion time alone but do not fully characterize gaze behavior or its temporal and spatial relationship to hand movement. Previous studies of eye–hand coordination in MS have demonstrated altered temporal coupling and shown that unstable gaze can exacerbate motor inaccuracies[10, 11]. However, these investigations have largely used controlled reaching or tracking paradigms, and relatively few studies have examined gaze during the 9-HPT itself [12]. Simultaneously measuring gaze, hand movement, and target location during this standardized task could connect laboratory observations of visuomotor dysfunction with clinically relevant dexterity performance.

This study characterized eye–hand coordination during the peg-placement component of the 9-HPT using simultaneous eye tracking, markerless hand tracking, and object detection. Gaze, hand, and target-hole positions were mapped frame by frame to pegboard coordinates, enabling each placement to be divided into transport and manipulation phases. The manipulation phase was defined as the interval from hand arrival at the target hole to completion of peg insertion. We examined phase durations, gaze behavior, eye–hand movement onset asynchrony, and the accuracy and variability of gaze and hand positioning, as well as their associations with clinical measures. We hypothesized that longer peg-placement times in PwMS would be concentrated in the terminal manipulation phase, reflecting the greater demands of precise alignment and insertion.

# 2 Methods

### *2.1 Participants*

We recruited a total of 18 participants, comprising 9 people with relapsing-remitting multiple sclerosis and 9 healthy controls. All PwMS had disease duration ranging from 3 to 23 years at the time of testing. Healthy controls had no history of neurological disorders, visual impairment, or upper extremity dysfunction. Visual function was characterized in both groups with low-contrast Sloan letter acuity and Randot stereoacuity, and in PwMS additionally with the King-Devick test and the Convergence Insufficiency Symptom Survey. Demographic characteristics of both groups are summarized in Table 1. The study was approved by the Kessler Foundation Institutional Review Board. Visual and oculomotor history was recorded by self-reporting on a background screening questionnaire. Some PwMS reported a prior or intermittent visual disturbance, such as history of optic neuritis or diplopia, but none had a condition that affected eligibility, all completed the 9-HPT with both hands.

| **Characteristic** | **PwMS (n=9)** | **Controls (n=9)** |
|---|---|---|
| **Age, years, mean ± SD (range)** | 59.1 ± 7.0 (46-65) | 37.3 ± 14.3 (24-57) |
| **Sex, women/men, n** | 9/0 | 7/3 |
| **MS phenotype, RRMS, n** | 9 | N/A |
| **Disease duration, years, mean ± SD (range)** | 12.1 ± 6.2 (3-23) | N/A |
| **Ambulation Index walk time, s** | 7.1 ± 2.4 | Not tested |
| **King-Devick time, s** | 57.7 ± 9.8 | Not tested |
| **Low-contrast letter acuity, 2.5%** | 42.1 ± 6.9 | 53.1 ± 10.5 |
| **Randot stereoacuity, median (range)** | 70 (20-400) | 30 (20-50) |

**Table 1.** Participant characteristics. SD, standard deviation; RRMS, relapsing-remitting multiple sclerosis. Low-contrast letter acuity is the number of letters read on the 2.5% Sloan chart. N/A, not applicable

### 2.2 Apparatus and Equipment

The experimental setup consisted of three integrated systems for multimodal data acquisition: eye tracking, object detection and segmentation and markerless hand tracking.

**Eye Tracking.** Gaze position was recorded using Tobii Pro Glasses 3 at 100 Hz. Before each session, the eye tracker was calibrated using the single-target calibration card, positioned approximately 0.5-1 m from the participant.

**Object Detection and Segmentation.** YOLO26x segmentation model was trained on a custom dataset of 2,500 augmented images collected during pilot testing. The model was used to detect and segment task-relevant objects, such as individual pegs, pegboard holes, the collection bin, the participant's hands, and the pegboard.

**Hand Tracking.** Hand position was estimated using WiLoR [13], a transformer-based model for three-dimensional hand localization and reconstruction. Video data were acquired using the integrated scene camera of the Tobii Pro Glasses 3 at 25 Hz, enabling frame-by-frame tracking of hand position throughout each trial.

### 2.3 Nine-Hole Peg Test Procedure

The 9-HPT is a standardized measure of upper extremity fine motor dexterity widely used in MS clinical trials and research. The apparatus consists of a shallow container holding nine small pegs and a board with nine empty holes arranged in a 3×3 grid. Participants are instructed to pick up the pegs one at a time, place them into the holes as quickly as possible, and then remove them one at a time and return them to the container.

Participants were seated at a table with the pegboard positioned at a comfortable reaching distance. Following eye-tracker calibration, participants completed two trials with their dominant hand, followed by two trials with their non-dominant hand. The total 9-HPT completion time, including both peg placement and removal, was recorded for each trial, and the mean completion time was calculated for each hand. For the multimodal analysis, each trial was further divided into peg-placement and peg-removal phases, with the duration of each phase calculated separately. The present study focused on the peg-placement phase, which was further segmented into individual peg placements to characterize phase-specific eye–hand coordination.

Throughout each trial, hand position, gaze position, and task-relevant object locations were recorded continuously and synchronized using timestamps. Clinical measures included the examiner-timed 9-HPT, the Ambulation Index walk time, low-contrast letter acuity (2.5% Sloan chart) and the King-Devick test.

## 2.4 Data Processing

### 2.4.1 Hand Kinematic processing:

Hand position was estimated using WiLoR [13]. To distinguish participants' hands from those of the experimenter, detected hands were evaluated based on wrist landmark orientation relative to other hand landmarks in the participant's egocentric view. The active hand was identified based on the hand used during each trial and its predominant movement, while detections corresponding to the experimenter's hands or inactive hands were excluded.

When WiLoR failed to detect the participant's hand, YOLO-generated hand bounding boxes were used to define regions of interest (ROIs), within which hand detection was repeated at an increased image resolution.

To compensate for participant head movement, a Lucas–Kanade sparse optical flow tracker was used to estimate camera displacement from a user-defined stable background region. The resulting displacement series was smoothed and subsequently subtracted from the hand landmark coordinates to reduce the influence of camera motion on the estimated hand trajectories.

### 2.4.2 Eye tracking processing:

Gaze data from the Tobii Pro Glasses 3 were exported as tab-separated files at 100 Hz. Recordings were trimmed to trial boundaries using event markers embedded in the data files. Saccade and fixation data were exported using the Tobii Pro Labs attention filter. Saccade amplitude was converted from scene-camera pixels to degrees of visual angle with a pinhole model of the scene camera (1920x1080px, 95 deg x 63 deg field of view, focal length 880px).

### 2.4.3 Object detection processing:

Eye tracker's world view was used to maintain the pixel-based coordinate system for all trackers. YOLO was run in tracking mode using BoT-SORT algorithm for tracking each object's location across frames and their state tracking. Object-based ROIs were defined around the pegboard holes and collection bin to support the identification of peg-placement and removal events. When the participant's hand entered these regions, the corresponding object-tracking information was temporarily held to account for hand-related occlusion during peg manipulation.

### 2.4.4 Mapping eye-hand data to board coordinates

Though the physical distance between the holes is the same (32 mm center to center, measured on the physical board), pixel distances appear different in the head-mounted scene camera image, from near hole row to far hole row. Pixel distances therefore cannot be converted with a single scale. A projective transformation (homography) was used to map eye-hand positions from the scene camera image to pegboard coordinates. Because participants view the pegboard from different and continually changing distances, the board's position in the image was estimated in every frame. The detected empty and filled holes were used to build the 3 x 3 lattice. Each hole was represented by the midpoint of the bottom edge of its bounding box, because a seated peg makes the box taller and shifts its centroid upward. This converted the eye and hand measures to board units, which account for perspective, viewing distance and head tilt.

The lattice was estimated in one of the three ways depending on holes and filled holes visibility in a frame. (1) when all the nine holes or filled holes were visible distance between bottom edge points of neighboring holes and filled holes were used to label each hole's row and column, and a homography was solved (2) When six to eight are visible, the lattice from the nearest solved frames was shifted by the head motion between two frames, measured by the optical-flow tracker (section 2.4.1), and were fitted to the detected holes ( a homography with eight holes, an affine fit with six or seven holes or filled holes visible). (3) when fewer than six holes were visible, usually because the hand occluded the board during placement, the four corners of the lattice from the nearest solved frame were each shifted by the measured head motion. A homography solved from these corners gave all 9 estimated positions, which were aligned to the detected holes when at least two holes or filled holes were visible in the frame. Because the lattice points lie on the bottom edge of each hole, each point was then shifted to the hole centre by half the height of an empty hole's bounding box (about 4 mm on the board), projected into each frame. Distances measured in board units were multiplied by 32 mm to give millimeters so that positions can be compared across the hole rows.

### 2.4.5 Temporal Segmentation of the Nine-Hole Peg Test

The total 9-HPT was divided into two components: peg-placement duration and peg-removal duration. Peg-placement phase was defined as the time from the beginning of the trial until all nine pegs had been inserted into the pegboard. Peg-removal duration was defined as the time from the initiation of peg removal until the final peg was returned to the collection bin. The 9-HPT completion time reports total time taken to finish these both components in a trial.

The present analysis focused on the peg-placement phase. Each trial was segmented into nine individual peg placements, and each placement was further divided into two movement phases. (1) Object transport, in which the peg was carried from the collection bin toward the target hole, was measured from the hand's grasp of the peg in the bin (section 2.4.8) to the hand's arrival at the target hole (section 2.4.7). (2) Object manipulation, in which the peg was aligned and inserted, was measured from hand arrival to the frame in which the peg was seated. Hand arrival therefore marks the boundary between the two phases. (3) A third interval, the return, was measured from the frame in which the peg was seated to the grasp of the next peg in the bin (pegs 1–8). It comprises the hand's travel back to the bin and the time spent in the bin before the next grasp; because the grasp is the frame of minimum thumb speed inside the bin, it includes the dwell in the bin up to that frame.

The duration of each phase was calculated using the corresponding event timestamps. This segmentation enabled examination of whether differences in peg-placement time were associated with object transport, object manipulation, return or all three.

### 2.4.6 Event detection & Annotation

To categorize the eye-hand data events, horizontal positional data was used, and each frame data point was classified into transport phase or a stationary phase based on the absolute frame-to-frame x displacement of thumb tip which was smoothed with a 5-frame uniform filter. Frames with smoothed displacement of > 30 pixels/frame were labelled as transport and frames below this were labelled as stationary. For the eye movement traces, Tobii's saccade data was marked as transport phase and fixation data was used as stationary phase. Peg placements were anchored to increase in the count of filled holes detected by YOLO and were backtracked to the precise placement point by automated inverse peak detector. These events were cross-checked with the video to ensure correct marking of events and for the trials where this automatic event detection failed, interactive GUIs were built with eye-hand traces for each frame and were annotated manually during atypical movements.

### 2.4.7 Spatial accuracy

Hand reach error was defined as the signed horizontal error across the board (along the rows), positive toward the bin, between the thumb tip at the hand's arrival at the target hole, the end of object transport (section 2.4.5), and the center of the target hole. These were computed from the lattice on the frame of measurement (section 2.4.4) and multiplied by 32 mm. Hand arrival was identified as the first frame within the stationary interval at which the thumb tip's frame-to-frame x-displacement fell below 30 pixels/frame while within 40 pixels of the target center; this frame also starts the manipulation phase. Each annotation was then reviewed against the video frame and the hand trajectory.

Eye reach error was defined as the same signed horizontal error between the gaze position at the first fixation on the target hole and the center of the target hole. The first target fixation was identified manually by inspecting the gaze x-position trace and selecting the first fixation directed at the hole location; the gaze position in that frame was used.

Pegs whose frame of measurement had no eye or hand position, or no visible holes, because of poor data quality were excluded. One control participant was excluded from the spatial-accuracy analyses because the eye-tracker calibration had failed, leaving a constant offset of the recorded gaze; the spatial measures are therefore based on nine PwMS and eight controls. Two measures were derived from the signed errors: the mean signed error across pegs, which describes where the eye or hand landed relative to the hole center on average, and its standard deviation across each participant's pegs for each hand, which describes peg-to-peg endpoint variability. For statistical testing, the mean signed error was analyzed per peg and the standard deviation per trial (section 2.4.9).

### 2.4.8 Movement Onset Asynchrony

Movement onset asynchrony (MOA) was defined as the temporal difference between hand movement onset and eye movement onset during each peg-placement movement.

Eye movement onset was identified as the onset of the first saccade occurring during the bin-to-hole transport phase. Hand movement onset was identified as the first frame following the point of minimum hand movement within the collection bin, corresponding to the initiation of hand transport toward the target hole. All annotations were done semi-automatically, with manual annotation applied where thresholding failed during atypical movements. The grasp frame was reviewed against the video and adjusted where the detector had failed.

### 2.4.9 Statistical Analysis

Because each participant contributed two trials per hand and nine pegs per trial, group differences were tested with linear mixed-effects models (Y ~ Group + (1 | Participant), restricted maximum likelihood), fitted separately for each hand. The unit of observation was one peg for the within-reach timing and spatial measures, and one trial for completion time, fixation and saccade counts, saccade amplitude and the per-trial standard deviations of placement error. Degrees of freedom for the group contrast were obtained by the Satterthwaite approximation [14]. Group means are reported as PwMS vs controls. Associations with clinical measures were examined within PwMS ($n = 9$) with Spearman's rank correlation; for low-contrast acuity, available in both groups, a pooled partial correlation adjusted for group and age was also computed. No correction was applied for multiple testing. Analyses used MATLAB R2025b (fitlme) and Python 3.11 (SciPy, statsmodels).

# 3 Results

## 3.1 Participant characteristics

The cohort included nine participants with relapsing–remitting MS and nine healthy controls. All participants with MS were women, whereas controls included both women and men. The MS group was approximately 22 years older on average than the control group, with disease duration ranging from 3 to 23 years. Participant demographic characteristics are presented in Table 1. PwMS had mild walking disability (Ambulation Index walk time 7.1 +/- 2.4 s; two used a walking aid), lower low-contrast letter acuity than controls (2.5% Sloan chart, 42.1 +/- 6.9 vs 53.1 +/- 10.5 letters) and a mean King-Devick time of 57.7 +/- 9.8 s.

### 3.2 Global task performance

The 9-HPT completion time was longer in PwMS compared to controls for both dominant and non-dominant hands: For dominant hand 25.9 ± 5.6 vs 20.7 ± 3.0s with p = 0.025, Same trend was followed in the non-dominant hand trials 29.5 ± 9.7 vs 21.5 ± 3.8s with p = 0.034 (Fig. 1). The peg-placement phase alone, from the start of the trial to the seating of the ninth peg, took 18.5 ± 4.4 vs 14.2 ± 1.9s; p = 0.009 with the dominant hand and 22.2 ± 8.5 vs 14.6 ± 3.1 s; p = 0.019 with the non-dominant hand.

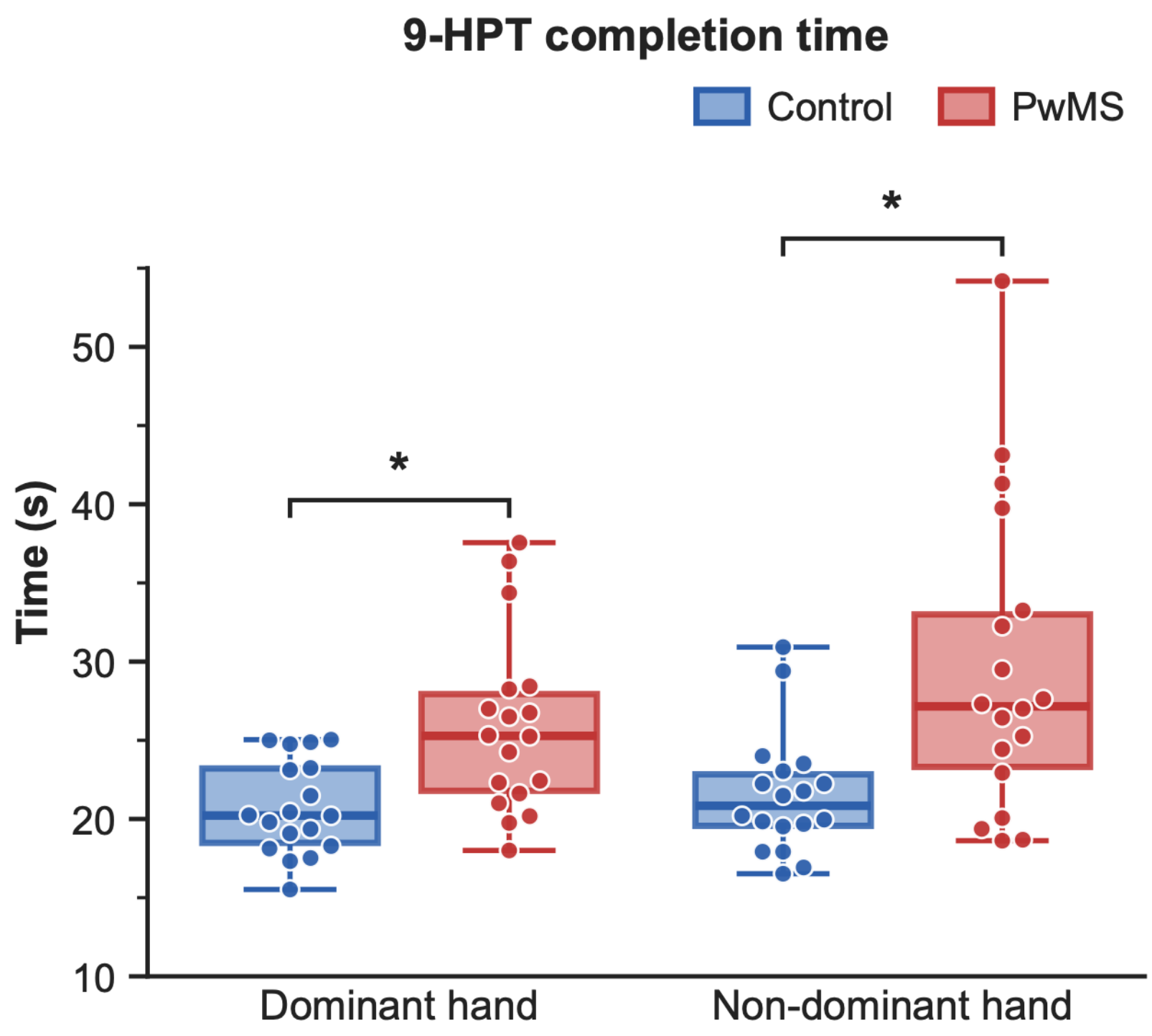

**Figure 1.** Overall, 9-HPT completion time. p value (*p < 0.05, **p < 0.01)

### 3.3 Representative Gaze Scan Paths and Hand Trajectories

Figure 2 shows one representative peg placement to a far-column hole from each group, drawn on the pegboard (upper panels, with fixations numbered in time order and joined by the scan path) and as the distance of the hand and the gaze from the pick-up position against time (lower panels). In both control examples (Fig. 2A, B) a single saccade carried gaze from the bin to the target hole, where it remained for 0.7 to 0.8 s while the hand travelled to the hole and inserted the peg, leaving for the bin only as the peg went in. In the PwMS examples gaze reached the target in stages: through two intermediate fixations short of the hole (Fig. 2C, fixations 2 and 3), or by returning from the target to the hand before fixating the target again (Fig. 2D, fixation 3). The hand paths of the PwMS examples were longer and less direct. These examples illustrate the phases of a reach; group differences were tested in the analyses that follow.

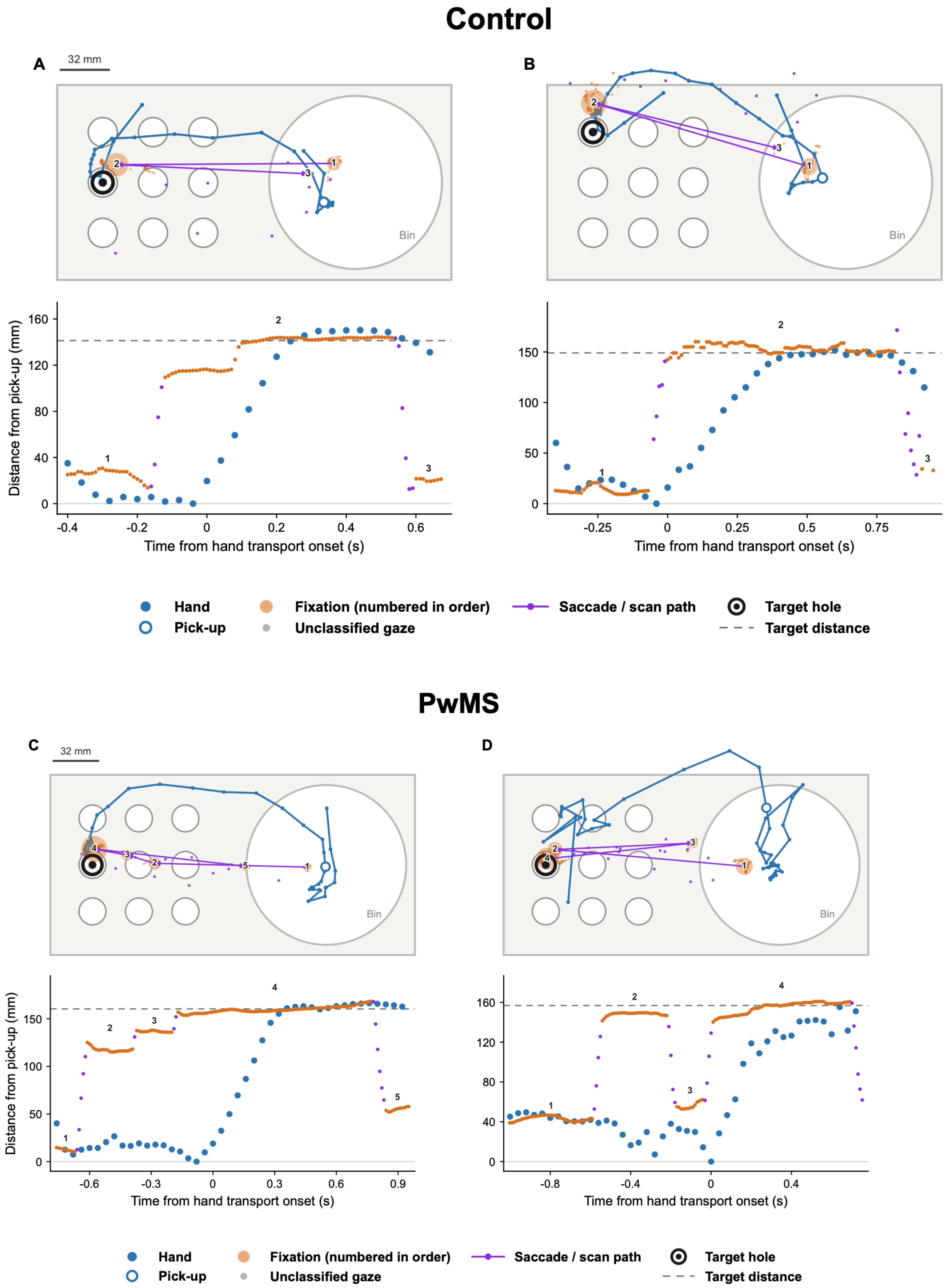

**Figure 2.** Representative peg placements to a far-column hole by two controls (A, B) and two people with MS (C, D). **Upper panels:** pegboard view (left-bin trials mirrored); blue line, thumb path (open circle, pick-up); orange discs (fixations; area proportional to duration); bull's-eye (target hole).

**Lower panels:** hand (blue) and gaze (orange) distance from pick-up over time from transport onset; dashed line represent the target-hole distance.

### 3.4 Gaze behavior

PwMS made more fixations per trial than controls with the dominant hand (35.9 vs 27.2, p = 0.010; Fig. 3a) and the non-dominant hand (44.5 vs 33.6, p = 0.019). The mean saccadic amplitudes were numerically larger in PwMS than controls for both the hands however neither difference reached statistical significance (dominant: 8.3 vs 7.9 deg, p = 0.61; non-dominant: 8.6 vs 8.0 deg, p = 0.48; Fig. 3b).

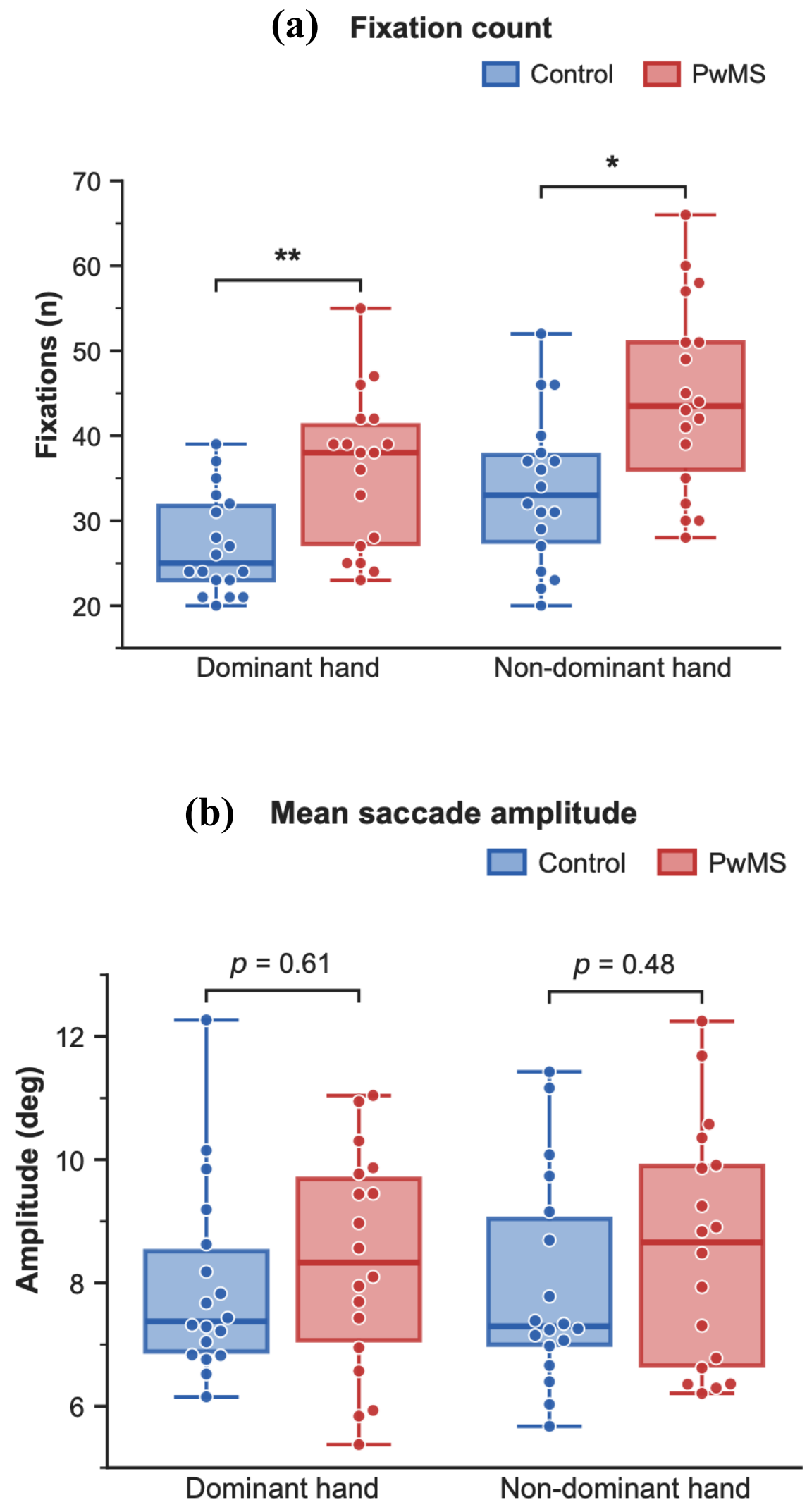

**Figure 3. (a)** Fixation count during peg-placement phase

**(b)** Mean saccade amplitude per trial (degrees of visual angle). Each point is one trial.

Gaze landing accuracy was tested as the signed horizontal (across-board) error between the gaze position at the first target fixation and the hole center, signed positive toward the bin (section 2.4.7), in nine PwMS and eight controls (one control was excluded for a failed eye-tracker calibration). The mean signed error did not differ between groups: gaze landed 11.2 vs 8.0 mm (dominant, p = 0.38) and 13.9 vs 11.5 mm (non-dominant, p = 0.58) on the bin side of the hole center (Fig. 4a). Its peg-to-peg variability was larger in PwMS with the dominant hand (SD 21.8 vs 16.1 mm, p = 0.027) but not with the non-dominant hand (21.2 vs 17.4 mm, p = 0.34; Fig. 4b).

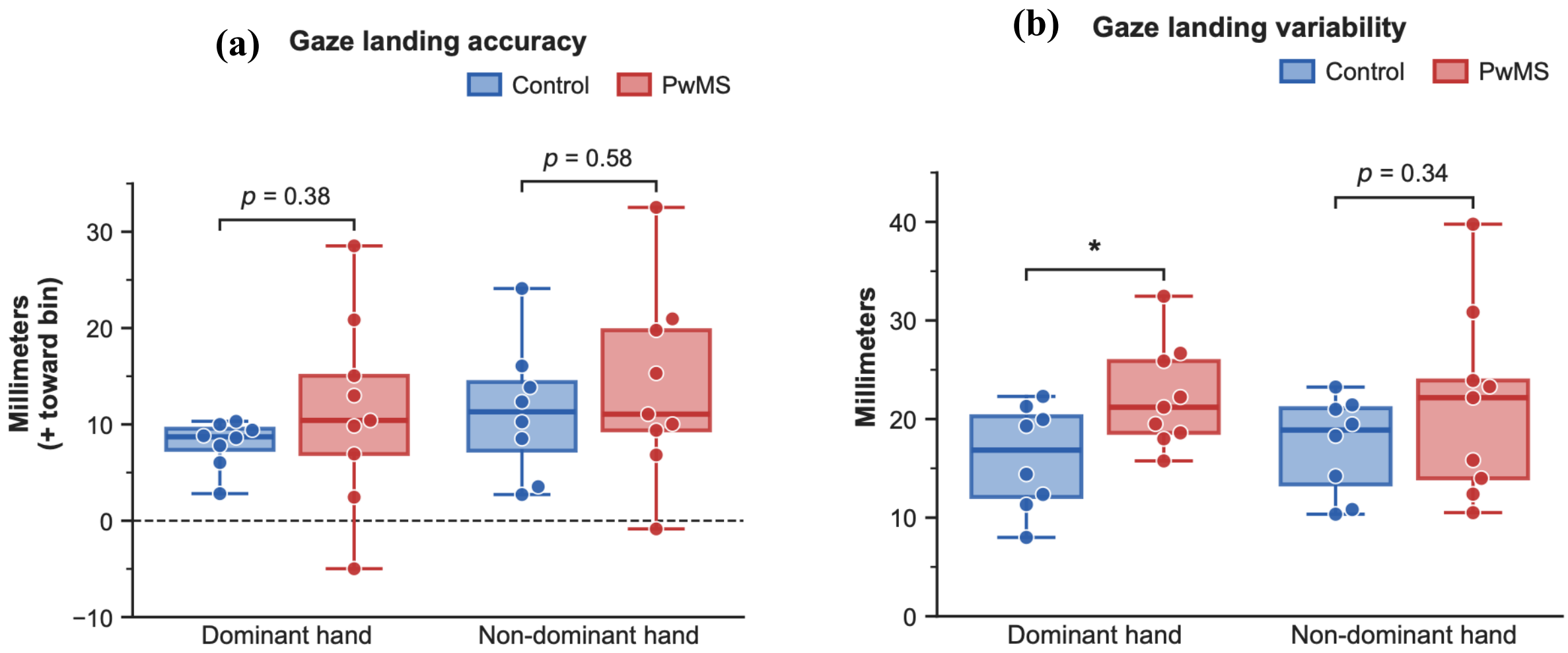

**Figure 4. (a)** Mean signed horizontal error between gaze at the first target fixation and the hole center, per participant (positive toward the bin).

**(b)** standard deviation of the same error across a participant's peg placements. Nine PwMS and eight controls (one control excluded for a failed eye-tracker calibration).

Gaze refinement time, from the first target fixation to the fixation closest to the hole center (Fig. 5), was likewise longer in PwMS (dominant: 632 vs 451ms, $p = 0.075$; non-dominant: 784 vs 522ms, $p = 0.023$).

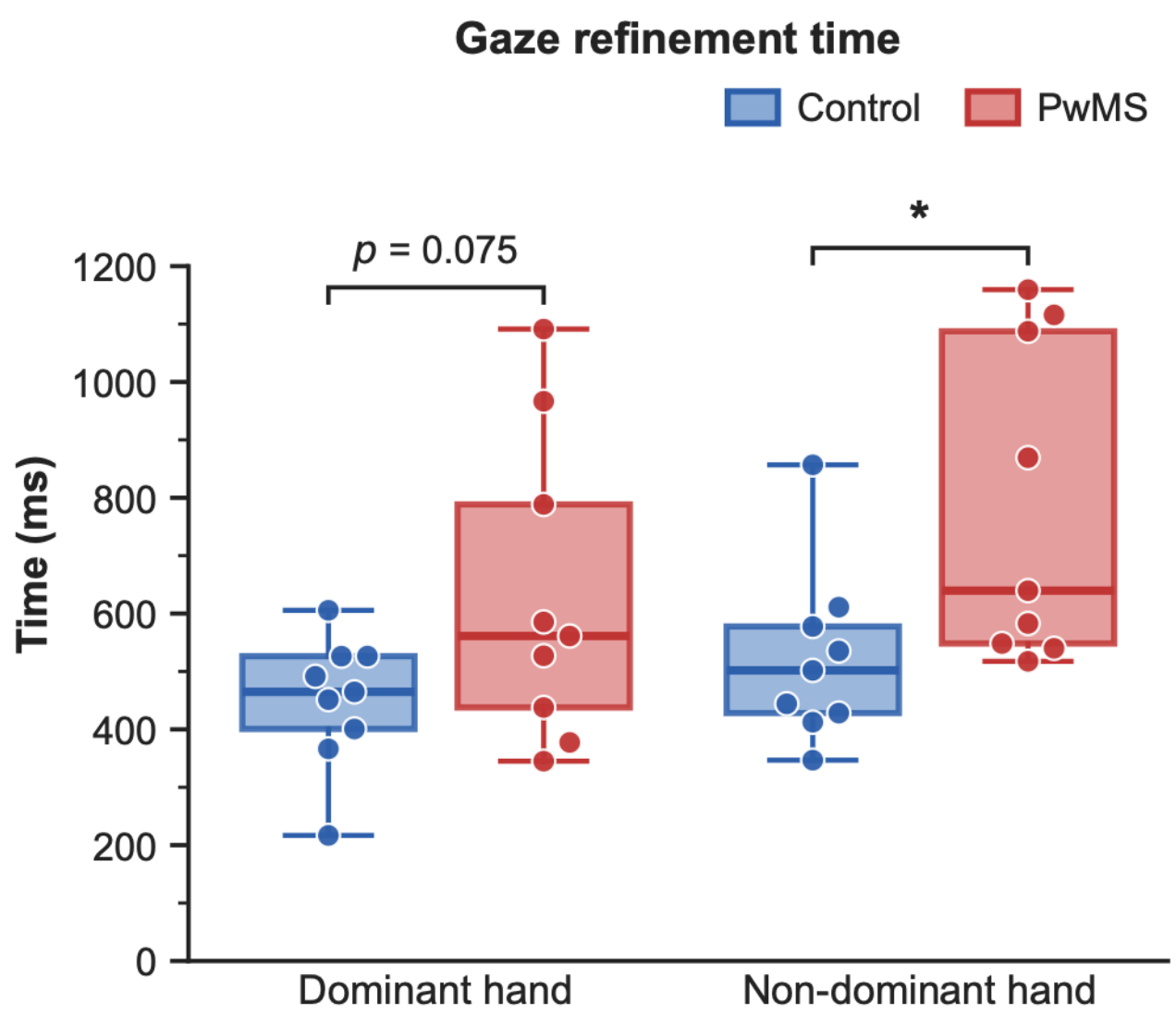

**Figure 5.** time from first target fixation to the fixation closest to the hole center. Each point is one participant.

### 3.5 Hand performance: object transport and manipulation

Object transport, from the grasp of the peg in the bin to the hand's arrival at the target hole (section 2.4.5), did not differ between groups with the dominant hand (429 vs 402 ms, p = 0.52) but was longer in PwMS with the non-dominant hand (563 vs 430 ms, p = 0.027; Fig. 6a). Object manipulation, from hand arrival at the target hole to completed insertion, was longer in PwMS with the dominant hand (544 vs 439 ms, p = 0.036) and numerically longer with the non-dominant hand (765 vs 477 ms, p = 0.105; Fig. 6b). The return, from the object manipulation completion to the grasp of the next peg, was longer in PwMS with both hands (dominant: 1028 vs 676 ms, p = 0.003; non-dominant: 1068 vs 700 ms, p = 0.021; Fig. 6c) and was the largest of the three phase differences, about 350 ms per peg.

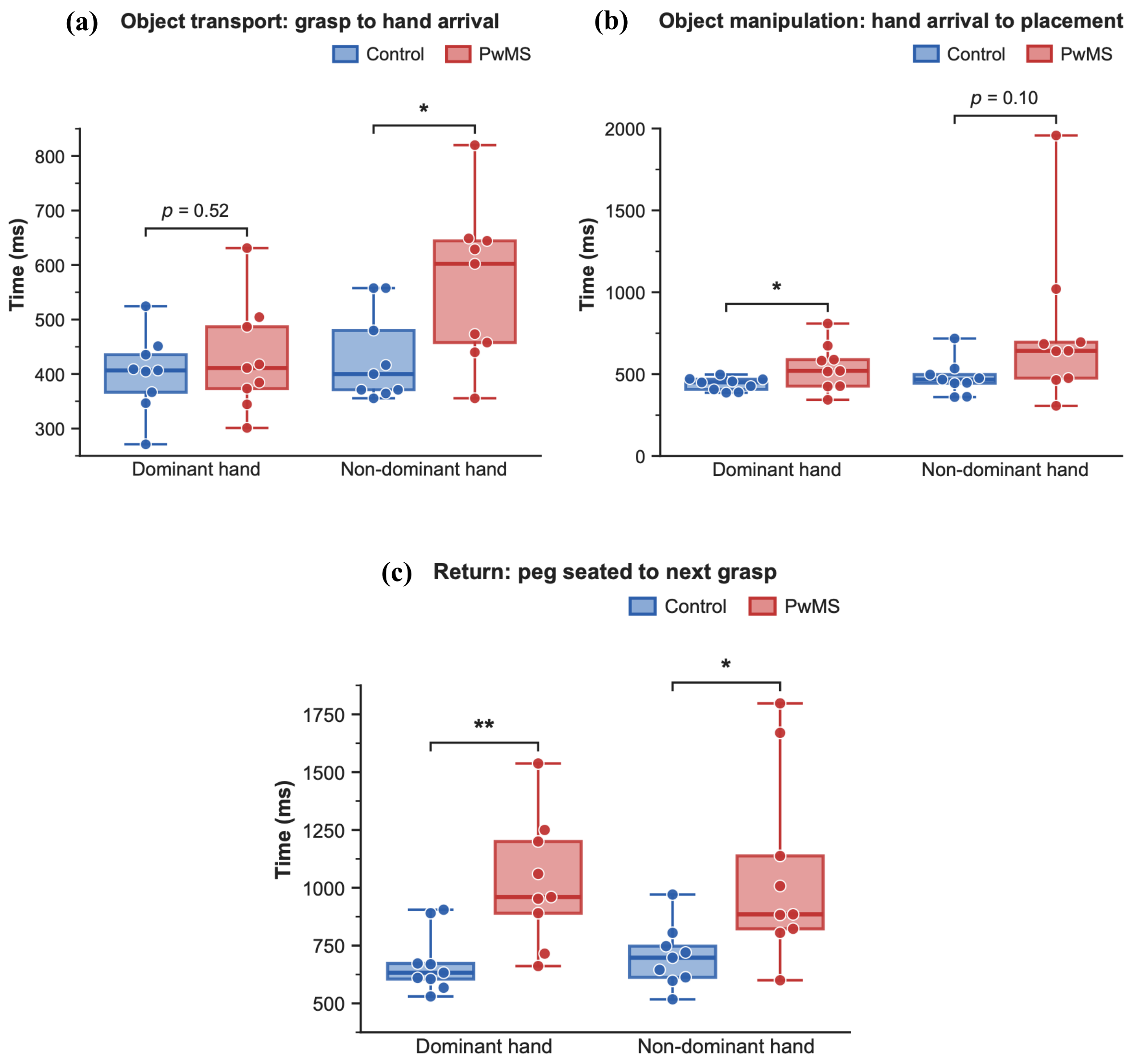

**Figure 6. (a)** time from the grasp of the peg in the bin to hand arrival at the target hole. **(b)** Object manipulation duration: from hand arrival to completed peg insertion. **(c)** from the seated peg to the grasp of the next peg (pegs 1–8). Each point is one participant

Hand positioning accuracy was examined as the signed horizontal error between the thumb tip at hand arrival and the hole center (section 2.4.7), in the same nine PwMS and eight controls. The mean signed error did not differ between groups: the thumb arrived 2.4 vs 1.3 mm (dominant, p = 0.75) and 3.1 vs 2.6 mm (non-dominant, p = 0.86) on the bin side of the hole center (Fig. 7a). Its peg-to-peg variability was numerically larger in PwMS with both hands, but neither difference was significant (dominant: SD 10.0 vs 8.1 mm, p = 0.25; non-dominant: 12.1 vs 9.0 mm, p = 0.16; Fig. 7b).

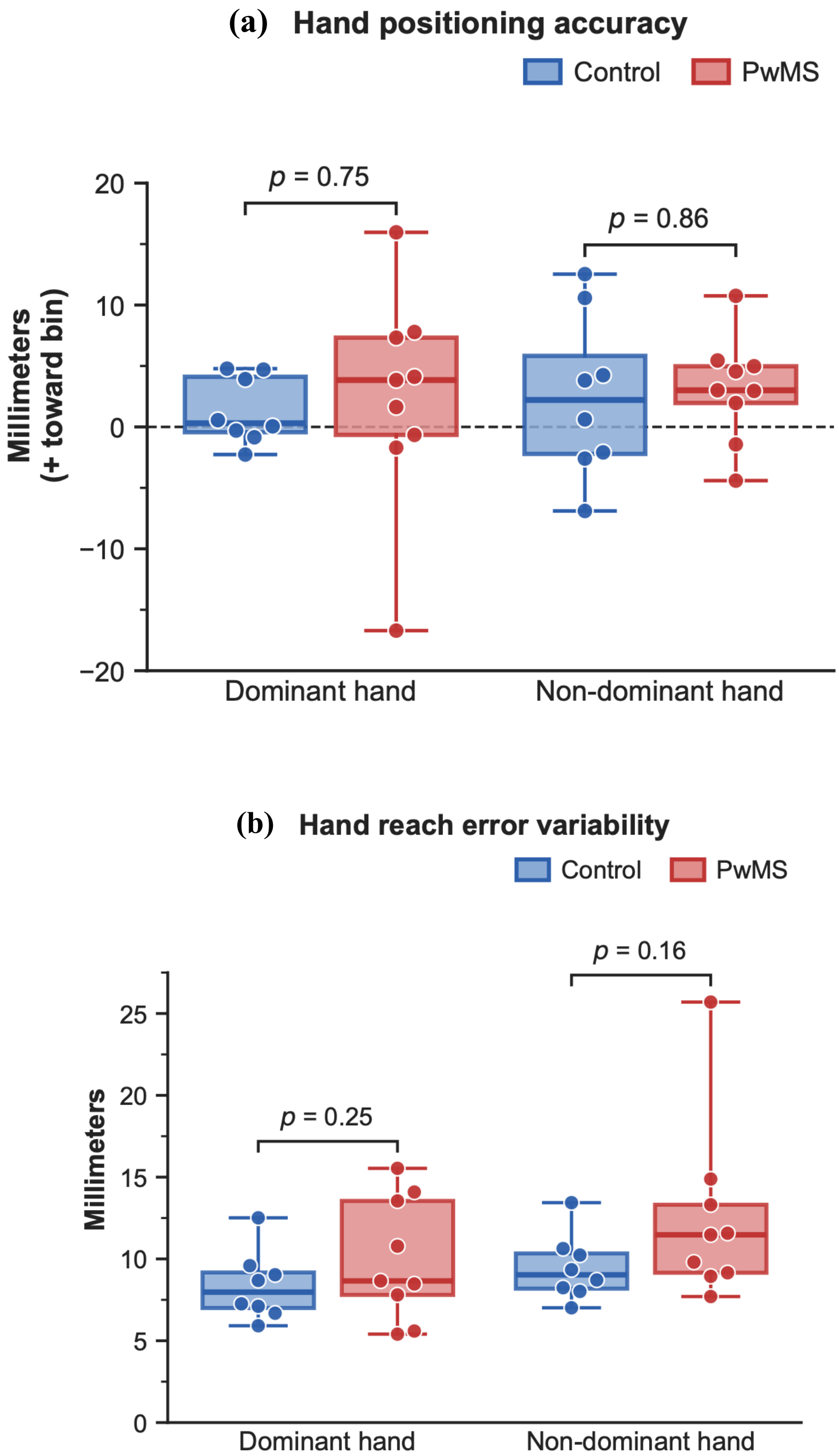

**Figure 7. (a)** mean signed horizontal error between the thumb tip at hand arrival and the hole center, per participant (positive toward the bin). **(b)** standard deviation of the same error across a participant's peg placements. Nine PwMS and eight controls, as in Fig. 4.

### 3.6 Eye-hand temporal coordination

Movement onset asynchrony, this is the mean temporal interval between the saccade onset and the hand movement onset towards the target hole (section 2.4.8). Positive MOA indicate that eye movement preceded hand movement. MOAs were positive across both groups for all trials confirming that the eye leads the hand consistently. The boxplots below show each subject's mean for each peg reach. MOA was longer in PwMS than in controls but not significant statistically. For dominant hand (271 vs 214ms, p = 0.085) and for non-dominant hand (257 vs 204ms, p = 0.22; Fig. 8).

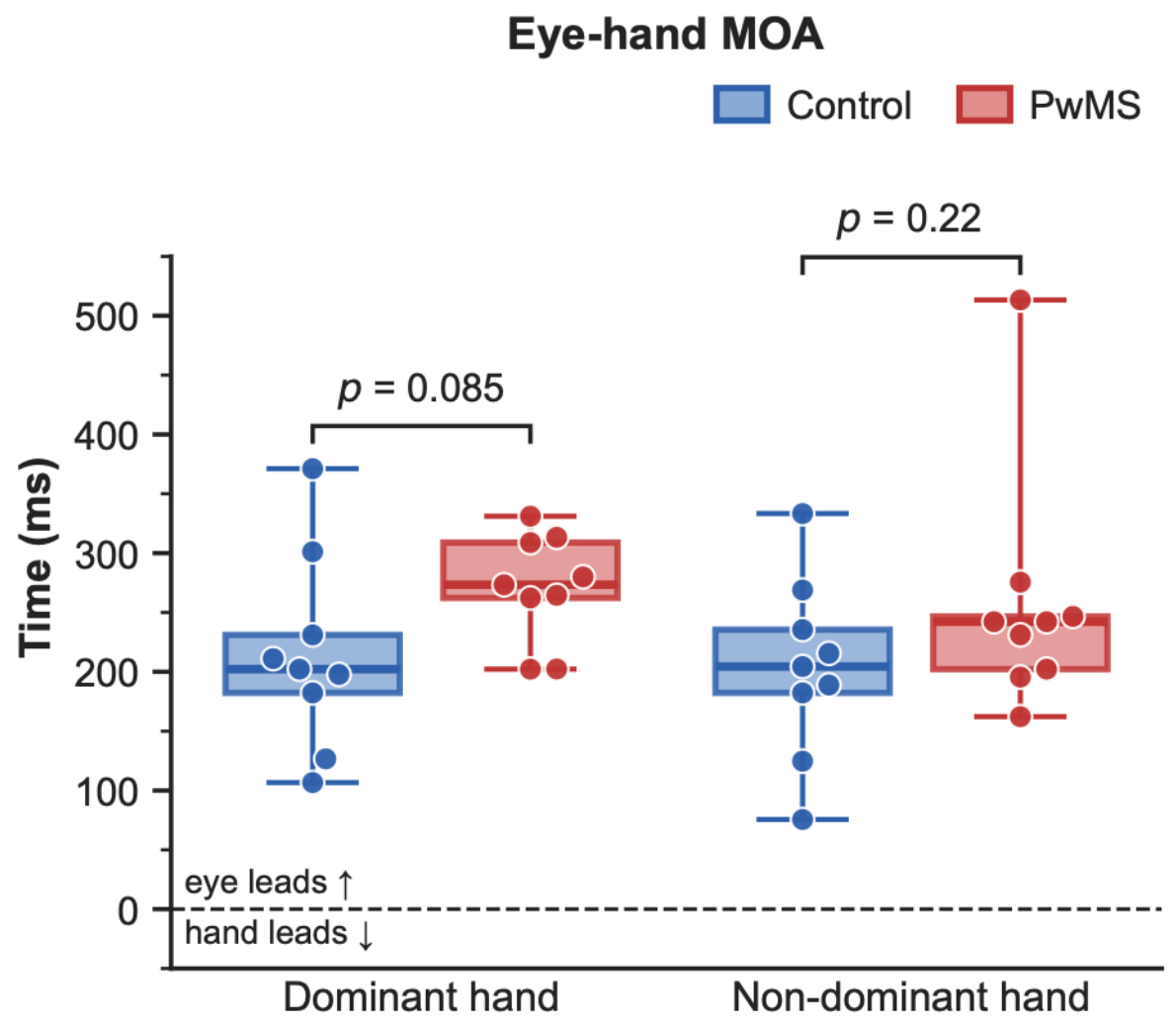

**Figure 8.** Movement onset asynchrony during peg placement. Positive values indicate that eye movement onset preceded hand movement onset.

The interval from the first target fixation to completed object manipulation, was longer in PwMS with the non-dominant hand (1402 vs 984ms, p=0.0495) and showed the same trend with the dominant hand (1078 vs 926ms, p=0.087; Fig.9).

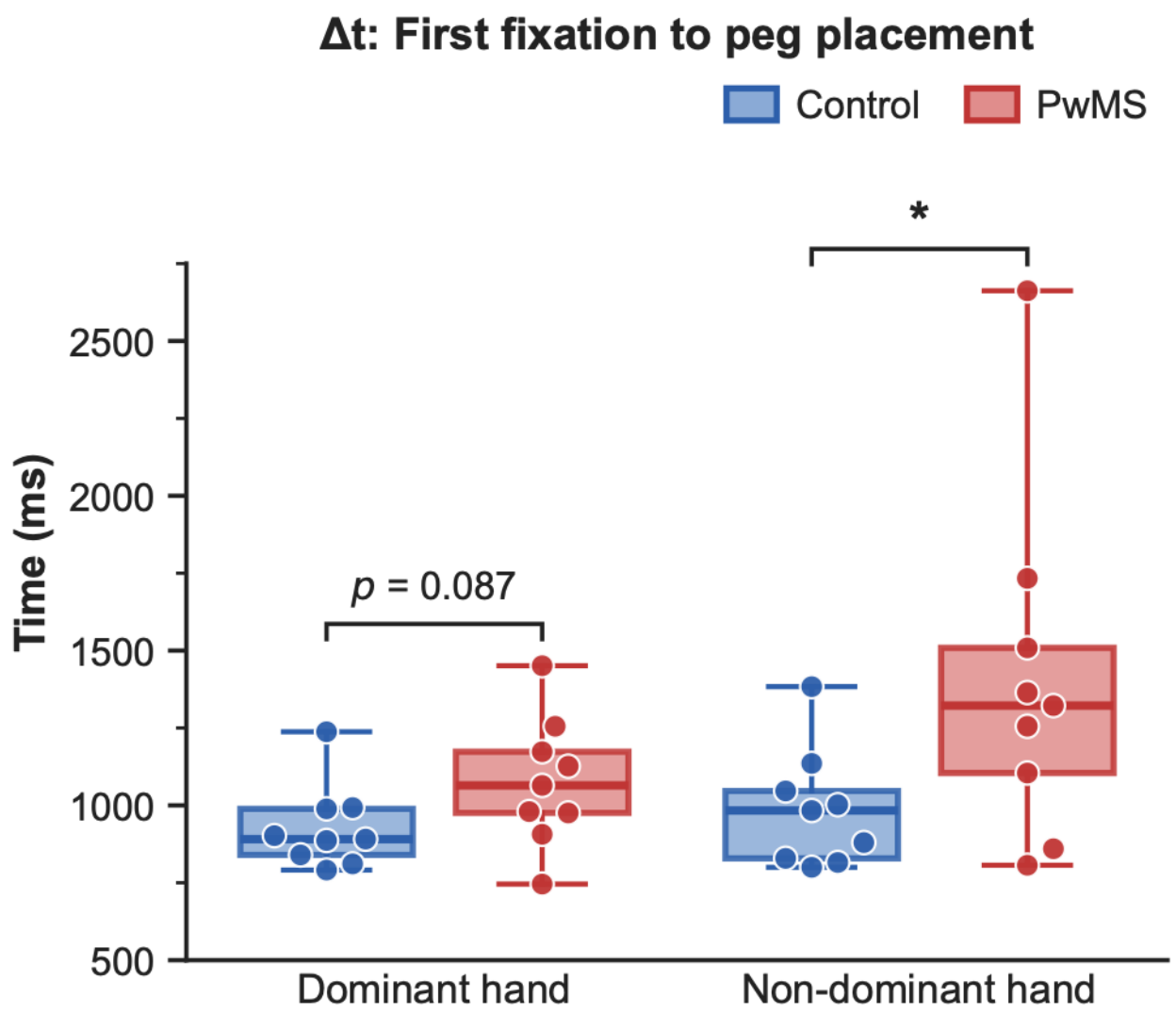

**Figure 9.** Interval from the first target fixation to completed object manipulation.

### 3.7 Associations with clinical measures

Within PwMS ($n = 9$), the Overall 9-HPT of each hand was associated with both the transport duration (dominant $\rho = 0.75$, $p = 0.024$; non-dominant $\rho = 0.87$, $p = 0.004$) and the manipulation duration of the same hand (dominant $\rho = 0.73$, $p = 0.032$; non-dominant $\rho = 0.96$, $p < 0.001$). The remaining associations use each participant's value over both hands during peg-placement phase. Ambulation Index walk time was associated with transport duration ($\rho = 0.82$, $p = 0.011$), manipulation duration ($\rho = 0.85$, $p = 0.006$) and peg-placement time ($\rho = 0.77$, $p = 0.021$). Low-contrast letter acuity (2.5% Sloan chart) was associated with peg-placement time in PwMS ($\rho = -0.77$, $p = 0.018$) and in controls ($\rho = -0.75$, $p = 0.036$, $n = 8$). King-Devick time was associated with peg-placement time ($\rho = 0.72$, $p = 0.037$) but not with saccade amplitude ($\rho = -0.35$) or fixation count ($\rho = 0.35$; all $p > 0.3$).

## 4 Discussion

Using simultaneous eye tracking, markerless hand tracking, and object detection, we characterized eye–hand coordination during the peg-placement component of the 9-HPT in nine PwMS and nine controls. The principal exploratory finding was that the largest share of the additional time in PwMS fell in the return from the seated peg to the grasp of the next peg, which was significantly longer with both hands, and, within peg placement, in object manipulation with the dominant hand. Manipulation durations were significantly longer in PwMS with the dominant hand and numerically longer with the non-dominant hand; transport duration was significantly longer only with the non-dominant hand. PwMS also made more fixations in all the trials and showed greater variability in gaze landing position during dominant-hand performance. Eye movement generally preceded hand movement in both groups, without a statistically significant difference in movement onset asynchrony. Together, these findings illustrate how separate measures of gaze, hand performance, and their temporal relationship can reveal features obscured by overall completion time.

### 4.1 Global task performance and representative movement patterns

Overall, 9-HPT completion time was longer in PwMS with both hands, consistent with previous reports of impaired manual dexterity in MS[8, 15]. Conventional scoring summarizes the time required to place and remove all nine pegs but does not identify which movement components contribute to slower performance. Separating peg placement into object transport and object manipulation extends previous efforts to instrument the test and characterize its component movements [9, 16, 17].

The representative scan paths and hand trajectories illustrate these components, including gaze shifts toward the target and hand movements preceding insertion. They help place the quantitative findings in the context of an individual action, while the group analyses establish whether these features differ between PwMS and controls.

### 4.2 Gaze behavior and target acquisition

PwMS made more fixations during peg placement with both hands, whereas saccadic amplitude did not differ significantly between groups. Greater fixation counts may partly reflect longer task duration rather than an increased rate of visual sampling. Nevertheless, this finding, together with greater dominant-hand variability in gaze landing position, raises the possibility that precise peg placement places additional demands on visual monitoring in MS [11].

Target acquisition was also slower in PwMS during non-dominant-hand performance. The interval from gaze departure at the bin to the first fixation at the target hole was significantly longer with the non-dominant hand, while the dominant-hand difference was nonsignificant. This interval describes gaze behavior after saccade onset and is distinct from saccadic latency. The recorded gaze trajectories permit identification of indirect paths, additional saccades, and intervening fixations that may prolong target acquisition. Although their individual contributions to this interval were not quantified in the present analysis, these features are plausible contributors to slower acquisition and are relevant to the saccadic abnormalities described in MS[18-20]. A more detailed analysis of the gaze sequence could clarify how these behaviors contribute to the observed timing differences.

The interval from initial target fixation to the fixation closest to the hole center was likewise significantly longer in PwMS with the non-dominant hand and numerically longer with the dominant hand. This measure may capture continued gaze adjustment during peg placement. However, because the closest fixation was identified within the interval ending at peg insertion, longer placement times provide a longer observation window. Its interpretation should therefore account for both the sequence of gaze adjustments and the duration of the accompanying manual action.

PwMS also showed greater peg-to-peg variability in horizontal gaze landing position with the dominant hand (21.8 vs. 16.1 mm). This finding indicates less consistent gaze positioning across placements, rather than necessarily a systematic error in target localization. Although compatible with reports of saccadic dysmetria and fixation instability in MS, variability across separate placements is distinct from instability within a fixation[19, 21]. Together, the fixation, timing, and spatial findings suggest altered visual sampling and target acquisition during peg placement, warranting further analysis of their relationship to subsequent hand adjustments.

### 4.3 Hand performance: object transport and manipulation

Object manipulation required an additional 105ms per peg in PwMS with the dominant hand and 287ms with the non-dominant hand. The dominant-hand difference was statistically significant; the non-dominant-hand difference was not. Transport did not differ with the dominant hand but was 133ms longer per peg with the non-dominant hand. The largest difference was the return, from the completion of object manipulation to the grasp of the next peg, which took about 350ms longer per peg in PwMS with both hands. The return duration is not a simple transport interval. It includes the search and grasp of the next peg in the bin, a visually guided precision task like insertion. This pattern supports the hypothesis that difficulty may be concentrated in the precise alignment and insertion required after the peg reaches the target, although the small sample limits conclusions about the relative contributions of the two phases.

Manipulation requires fine fingertip control under visual and tactile guidance. Longer manipulation duration and return durations could reflect additional positioning, gripping or corrective adjustments, consistent with previous reports of terminal aiming inaccuracies, altered visually guided reaching, and reduced movement smoothness in MS[6, 7, 22]. The recorded hand trajectories provide a basis for examining these adjustments, although their frequency and contribution to manipulation duration were not quantified here. Greater reliance on sensory feedback remains a plausible interpretation that would require further investigation.

Hand positioning variability was numerically greater in PwMS with both hands, but neither difference reached statistical significance. Thus, the clearest hand-performance findings were the longer return with both hands and the prolonged dominant-hand manipulation duration, rather than a demonstrated group difference in spatial consistency.

### 4.4 Eye–hand temporal coordination

MOA was positive in both groups, indicating that eye movement generally began before hand transport. Mean MOA was numerically longer in PwMS with the dominant hand (271 vs. 214 ms) and non-dominant hand (257 vs. 204 ms), but neither group difference was statistically significant. MOA describes the relative onset of eye and hand movements; it does not establish when gaze stabilized at the target or, by itself, identify delayed hand initiation.

Prior work by Rizzo, Hudson, and colleagues provides a useful framework for interpreting these relationships [23, 24]. In chronic middle cerebral artery stroke, a controlled saccade-to-reach paradigm demonstrated greater separation between eye and hand movement onsets, accompanied by additional corrective saccades and larger spatial errors [23]. That study showed that eye movement can precede hand movement while their temporal coordination remains disrupted. In the same population, providing augmented feedback about eye and hand endpoint errors after each reach improved eye–hand coordination, indicating that such coordination deficits are modifiable [25]. In the present MS cohort, however, the nonsignificant MOA differences do not establish comparable temporal decoupling. Differences in the tasks and event definitions also preclude direct comparison of MOA magnitudes across studies.

The interval from target fixation onset to completed peg insertion was significantly longer in PwMS with the non-dominant hand and numerically longer with the dominant hand. This interval spans both the remaining target approach and object manipulation, distinguishing it from movement onset asynchrony. Together, these measures suggest that examining coordination beyond movement initiation may be particularly informative during a task requiring precise object placement.

### 4.5 Associations with clinical measures and clinical relevance

Within PwMS, manipulation duration and transport duration were strongly associated with conventional 9-HPT completion time for both hands. These findings support the relevance of the manipulation and transport phase to overall dexterity performance. However, these contributes to total completion time, so the association does not establish independent or incremental clinical value.

Longer Ambulation Index walking time was associated with longer manipulation duration and trial time, suggesting that slower manual performance may accompany broader functional impairment. Poorer low-contrast letter acuity was also associated with longer trial time in both groups, with the relationship remaining after adjustment for group and age. These findings support further examination of visual function as a contributor to performance during visually guided manual tasks.

King–Devick time was associated with trial time but not with fixation count, saccade count, or saccadic amplitude. Because King–Devick performance involves visual scanning, attention, and rapid number naming, this association should not be attributed solely to saccadic function. Collectively, the clinical associations suggest that peg-test performance reflects multiple interacting processes, although the small sample and exploratory analyses warrant caution.

The multimodal framework preserves the standard peg-test task while providing additional information about gaze behavior, hand positioning, and movement timing. Previous robotic assessments have similarly shown that component measures can distinguish PwMS with and without self-reported hand difficulties when total 9-HPT completion time does not [26]. Further work is needed to establish whether the present measures add clinically useful information, demonstrate adequate reliability, and remain practical when calibration, tracking, and annotation requirements are considered.

**4.6 Limitations**

Several limitations affect interpretation. Participants with MS were approximately 22 years older than controls, and all were women, whereas controls included women and men. Age and sex differences may have contributed to the observed performance differences. Older adults may rely more heavily on visual feedback during precise manual actions and show altered gaze and hand behavior on pegboard tasks [12, 27]. Larger cohorts matched for age and sex are therefore needed to distinguish MS-related effects from demographic influences.

The sample was small, comprising nine participants per group, and multiple outcomes were tested without correction for multiple comparisons. Both group differences and clinical correlations should be considered exploratory. Nonsignificant findings do not establish equivalence, and the results require replication before conclusions can be drawn about phase-specific impairment or clinical utility.

Intermittent hand occlusion and reduced object-detection accuracy affected tracking continuity in some participants. Manual review and annotation were required when automated event detection failed, and placements with insufficient spatial data were excluded. These factors may affect reproducibility and introduce bias if missing data differ between groups or are more common during difficult movements. Reporting data retention and validating event detection against an independent reference would strengthen future studies.

Spatial analyses were limited to signed horizontal errors because vertical gaze measurements were less reliable. These measures do not capture full two-dimensional positioning accuracy, and a mean signed error can obscure opposing errors across placements. In addition, hand position was represented by the thumb tip rather than direct measurement of peg orientation or fingertip forces. The manipulation interval therefore identifies time spent between hand arrival and insertion without resolving the specific adjustments responsible for that time.

Hand tracking was limited to the scene camera's 25-Hz sampling rate, restricting the temporal precision of movement onset and phase-boundary estimates. Finally, longer task duration can influence fixation counts and the observation window for gaze-refinement measures. Future studies should evaluate duration-adjusted measures, analyze the gaze and hand adjustments within each phase, use higher-frame-rate hand recordings, and establish the reliability and validity of the derived outcomes.

## 5 Conclusion

Simultaneous eye tracking, markerless hand tracking, and object detection enabled characterization of gaze and hand behavior during the standard 9-HPT beyond overall completion time. In this exploratory cohort, PwMS had longer completion times, with the largest increase in the return from the seated peg to the grasp of the next peg with both hands, a significant increase in object manipulation with the dominant hand, and a smaller increase in object transport that was significant only with the non-dominant hand. Greater fixation counts and dominant-hand gaze landing variability further characterized performance, while MOA did not differ significantly between groups. These findings support further investigation of phase-specific eye–hand measures as complements to conventional dexterity assessment. Larger, demographically matched studies are needed to establish their reproducibility and clinical value.